# A VECTOR CONTROL AND DATA ACQUISITION SYSTEM FOR THE MULTICAVITY LLRF SYSTEM FOR CRYOMODULE1 AT FERMILAB *

P.Varghese#, B. Chase, B. Barnes, J. Branlard, E. Cullerton, P. Joireman, V. Tupikov
Fermi National Accelerator Laboratory, Batavia, IL, U.S.A.

*Abstract*

A LLRF control and data acquisition system for the 8-cavity cryomodule1 at the ILCTA has been implemented using three 33-channel ADC boards in a VXI mainframe. One card each is dedicated to the cavity probes for vector control , forward power and reverse power measurements. The system is scalable to 24 cavities or more with the commissioning of cryomodules 2 and 3 without additional hardware. The signal processing and vector control of the cavities is implemented in an FPGA and a high speed data acquisition system with up to 100 channels which stores data in external SDRAM memory. The system supports both pulsed and CW modes with a pulse rate of 5Hz. Acquired data is transferred between pulses to auxiliary systems such as the piezo controller through the VXI slot0 controller. The performance of the vector control system is evaluated and the design of the system is described.

## Introduction

The ILCTA (ILC Test Area) at Fermilab will consist of three 8-cavity cryomodules in a 24-cavity vector control configuration driven by a single klystron. Cryomodule1 has been installed and is currently undergoing cavity coupler conditioning. The LLRF system has been installed and is currently configured to process the 8-cavities in cryomodule1. Three Multi-cavity Field Control modules (MFC) are used to process the 75 channels of signals from the complete 24 cavity system[1]. The LLRF system is shown in Fig. 1. MFC1 is connected to the 24 cavity probe IF inputs and the reference IF input. MFC2 and MFC3 are similarly connected to the 24 forward power and 24 reverse power IF inputs respectively along with a reference IF input to each module. Downconversion from 1.3 GHz RF to the 13MHz IF is performed by ten 8/1 channel receiver/transmitter modules developed for the Fermilab SRF test facilities [2]. Vector summation and feedback control of the cavities is performed by MFC1. All modules have a data acquisition system that stores waveforms in external SDRAM memory, that can be accessed between pulses by the VXI slot0 module for display on GUI's such as Labview. The downconversion to base band and PI control is implemented in the FPGA. A cavity simulator connected to spare DAC outputs, allows the testing and evaluation of the closed loop feedback control algorithms. An automatic LO drift compensation scheme is provided by the onboard DSP which is also used for various computing tasks in the control system.

## Vector Control System

The 8 cavity probe signals for cryomodule 1 after downconversion to an IF frequency of 13MHz, are connected to an 8-channel , 12-bit ADC with serial LVDS outputs sampled at fs = 1313/21 MHz. The IF signals are downconverted to base band by 18-bit, 101 elements deep , NCO sine and cosine tables computed in the DSP. The fixed point 2.16 format table allows for magnitude adjustments from 0 to ~ 2 and arbitrary phase rotation. The downconverted I and Q signals are summed and filtered with a combination of a single stage CIC filter followed by a 3 tap FIR filter to remove the NCO mixing products. Set-point and feed-forward tables for the I and Q branches are 16-bit and

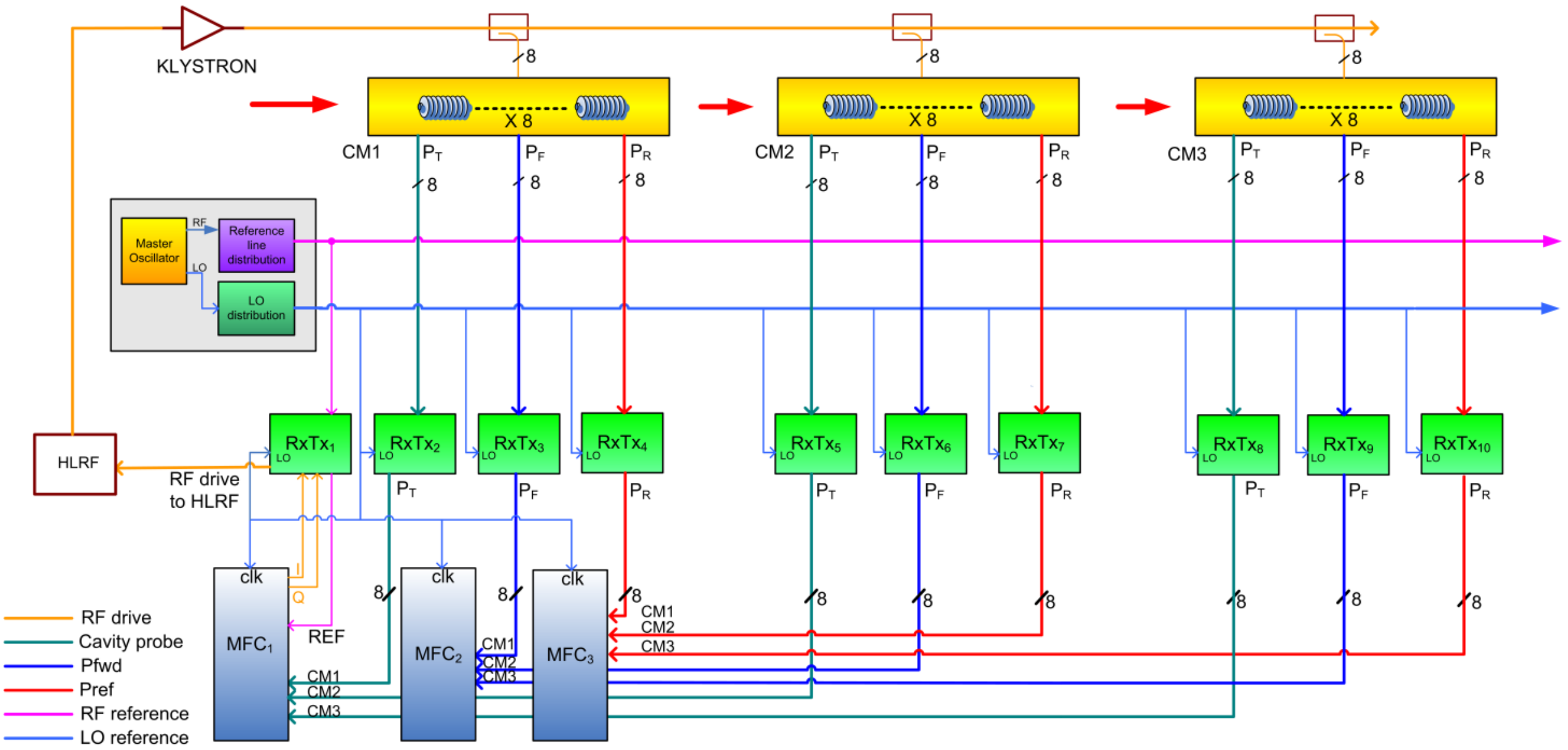

Figure 1: 24 - Cavity vector control system

*Work supported by Fermi Research Alliance LLC. Under DE-AC02-07CH11359 with the U.S. DOE

#varghese@fnal.gov

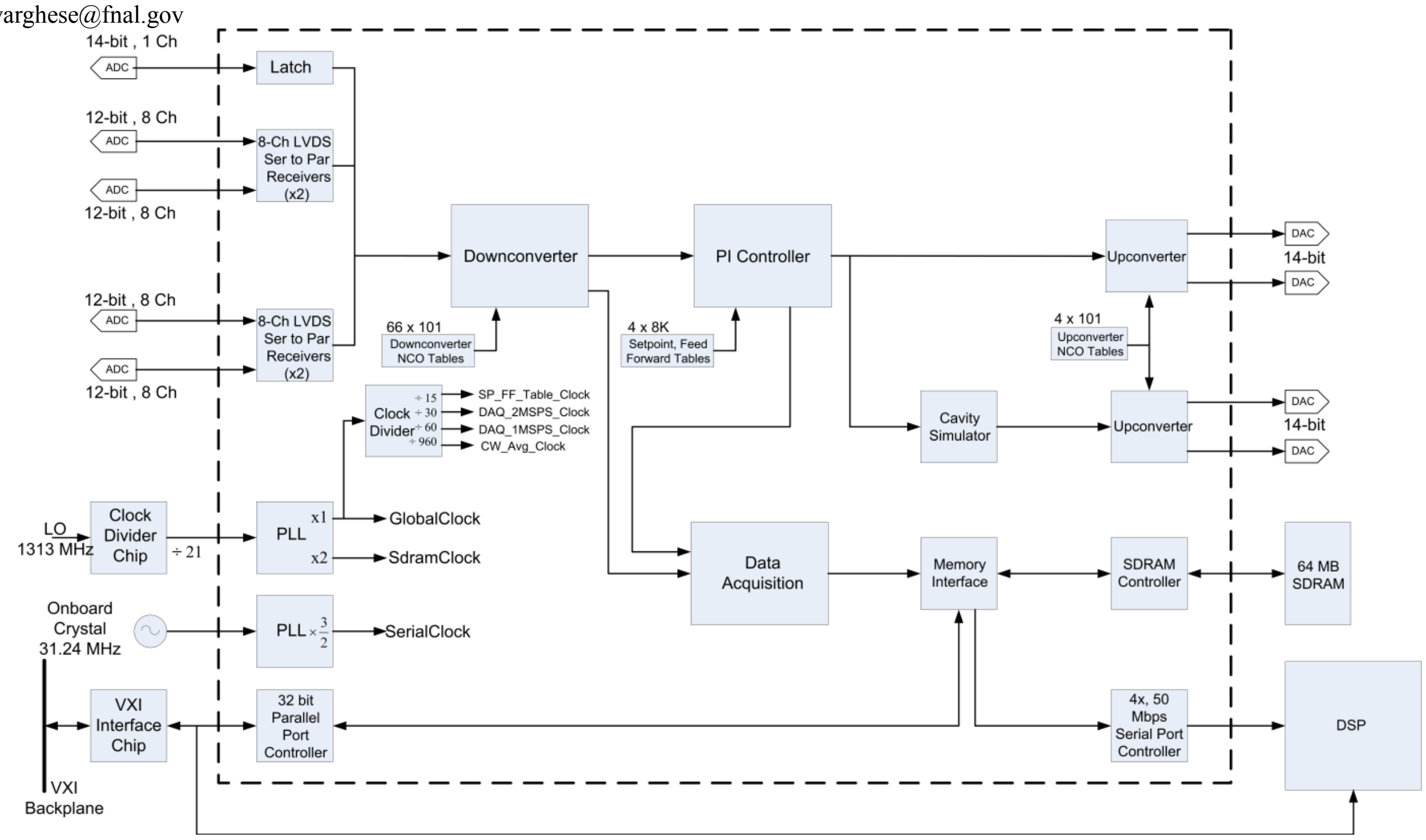

Figure 2: FPGA block diagram

8k words deep. The tables are computed in the VXI slot0 controller based on user parameters and are written to the FPGA memory in the interval between pulses. The error signals are fed to a PI controller with user settable gains and pole placement. Two notch filters, one fixed and one variable are used to remove the 8π/9 and 7π/9 mode frequencies. Complex upconversion to 13 MHz IF completes the signal processing chain before the I,Q outputs are sent to the 14-bit DACs which are connected to the transmitter inputs of the 8/1 channel receiver/transmitter. Magnitude correction and phase rotation for the upconveter NCO is provided in the same way as the downconverters. The controller output is also fed to a variable bandwidth base band cavity simulator for hardware simulations and closed loop testing. Both pulse mode and CW mode are sup-ported by the controller. The outline of the FPGA design is shown in Fig. 2.

## Data Acquisition System

The data acquisition system shown in Fig. 3 stores waveform data during the RF pulse which can be retrieved in the interval between pulses. Acquisition is initiated by the pulse trigger input. Data (16-bit) is latched on the same decimated clock edge from all channels and the latched data is transferred to a FIFO before a write to the external SDRAM. The decimation rate is nominally at fs/60 while the transfer to the FIFO is at a clock rate of 2fs or 3fs. This allows for a maximum of ~180 channels of data to be stored in an interleaved format in the SDRAM. For the 24 cavity system there are 25 ADC inputs (1 for reference) which produce 50 channels of I and Q data. The vector sum and 7 selected tap points along the signal processing chain contribute to 16 additional I and Q channels which are sampled at twice the rate, resulting in 32 additional channels to be stored in the SDRAM.

The waveform depth of ~2k words per channel covers the ~2ms duration of the RF pulse with a sampling rate of 1MS/s for the downconverted cavity probe inputs and a 2MS/s rate for the vector sum and signal tap points. A diagnostic mode is also provided to store any selected channel of the raw ADC inputs or DAC outputs at the full sampling rate for a maximum depth of 32 MS which is the size of the SDRAM (64 MB). For CW mode the averaged data for each channel is stored in a small block of dual port memory in the FPGA internal RAM.

## LO Drift Compensation

The phase stability for the LO distribution system in a vector control scheme is typically specified with a wider margin than the reference line distribution system. Any drift in the LO phase will appear as a measurement error for the cavity phase due to the fact that the mixing from RF to IF and the digital down-conversion from IF to base band frequencies does not affect the phase offsets which are transferred to the base-band signal. If the digital processing is performed with a clock derived from the LO, which is typically the case, there are further uncompensated error components appearing in both the measurement path and the drive or actuator path in the feedback loop. If the LO phase drift with respect to the reference can be measured, the corrections for the phase can be provided. A 64 sample subset of the reference and cavity probe I and Q data is sent to the DSP over the serial ports after each pulse. Any deviation from zero phase of the reference is corrected by rotating the NCO tables of both the reference and the cavity downconverters and the drive upconverter. The entire process takes ~2ms to complete and is scheduled at the end of the RF pulse before the waveforms are uploaded by the slot0 controller. The cavity probe I,Q data can be used in amplitude calibration or adaptive control schemes to improve the feedback controller's performance.

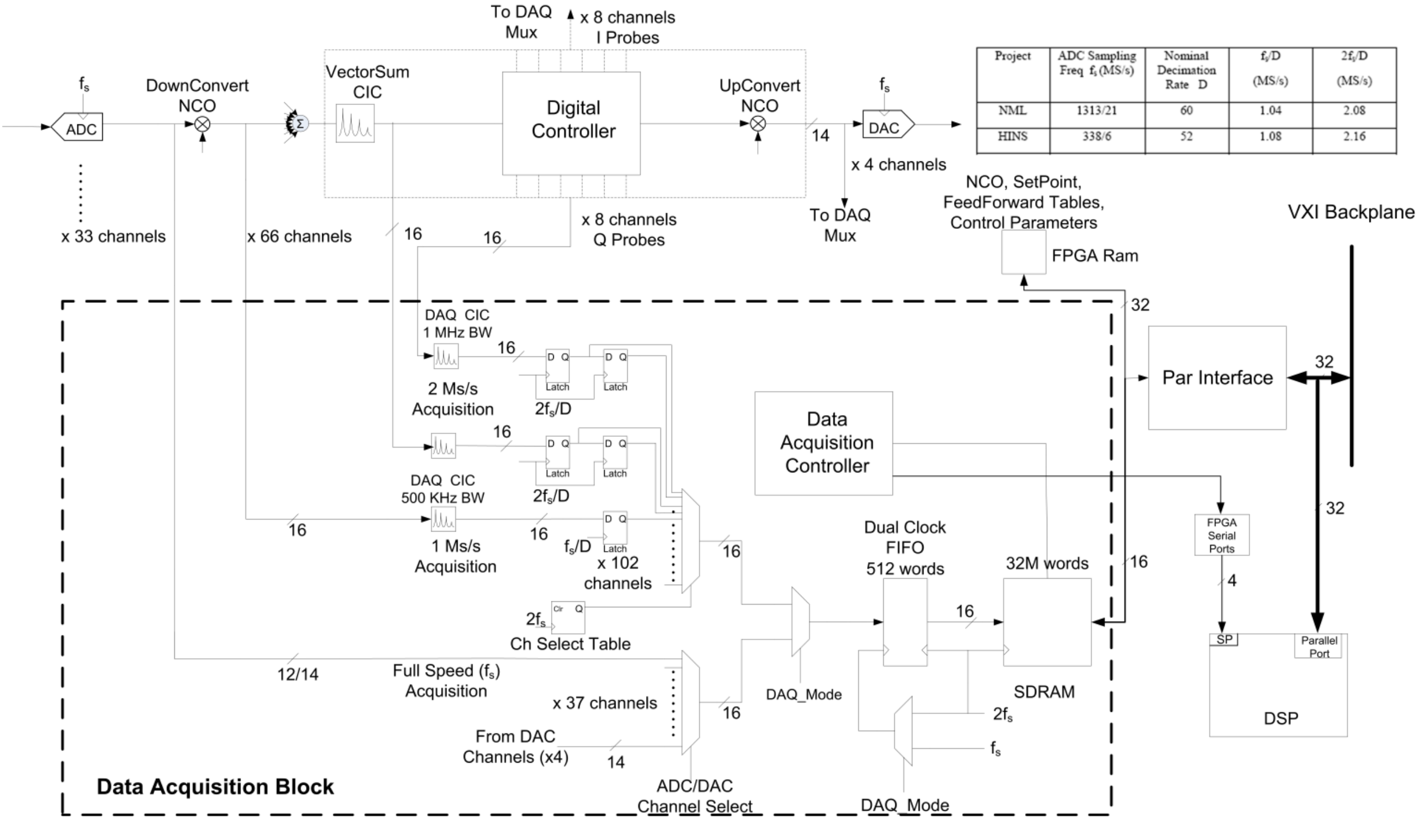

| Project | ADC Sampling Freq $f_s$ (MS/s) | Nominal Decimation Rate D | $f_s/D$ (MS/s) | $2f_s/D$ (MS/s) |
|---|---|---|---|---|
| NML | 1313/21 | 60 | 1.04 | 2.08 |
| HINS | 338/6 | 52 | 1.08 | 2.16 |

Figure 3: Data Acquisition System

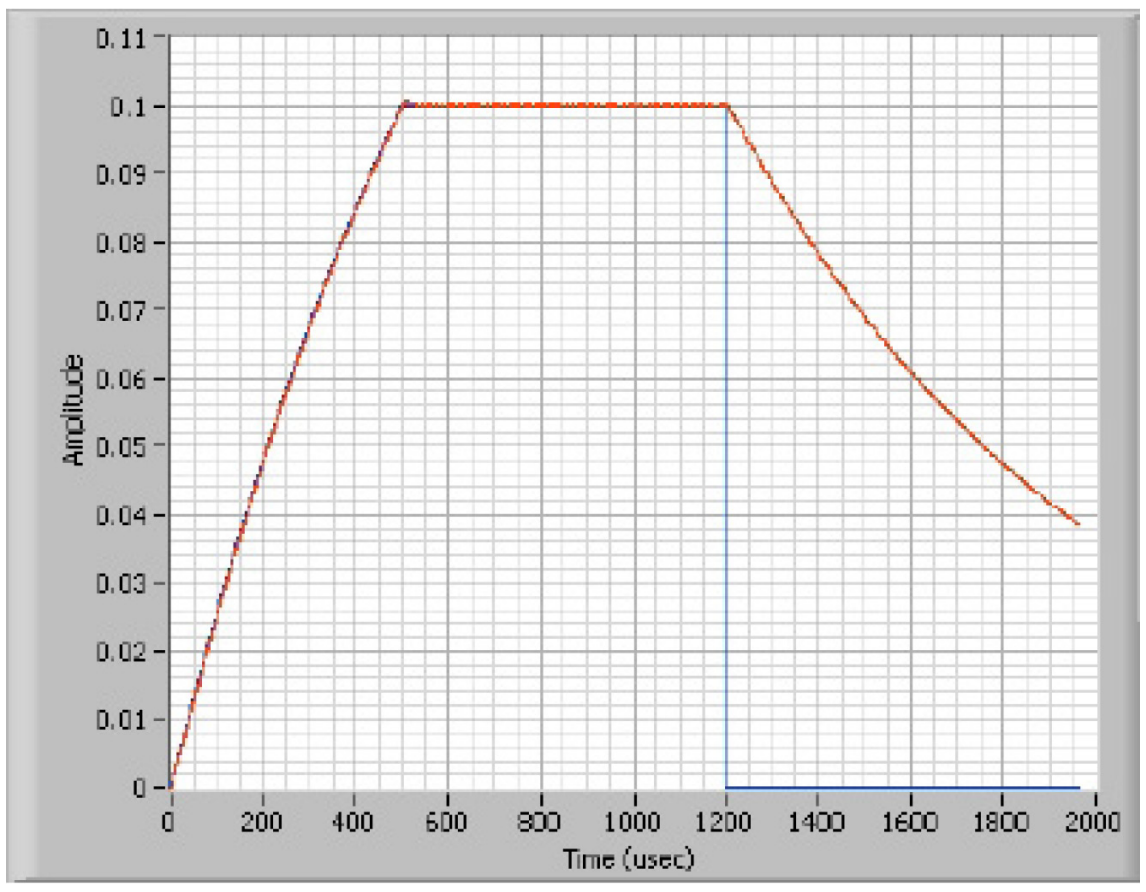

Figure 4: Vector Sum Amplitude

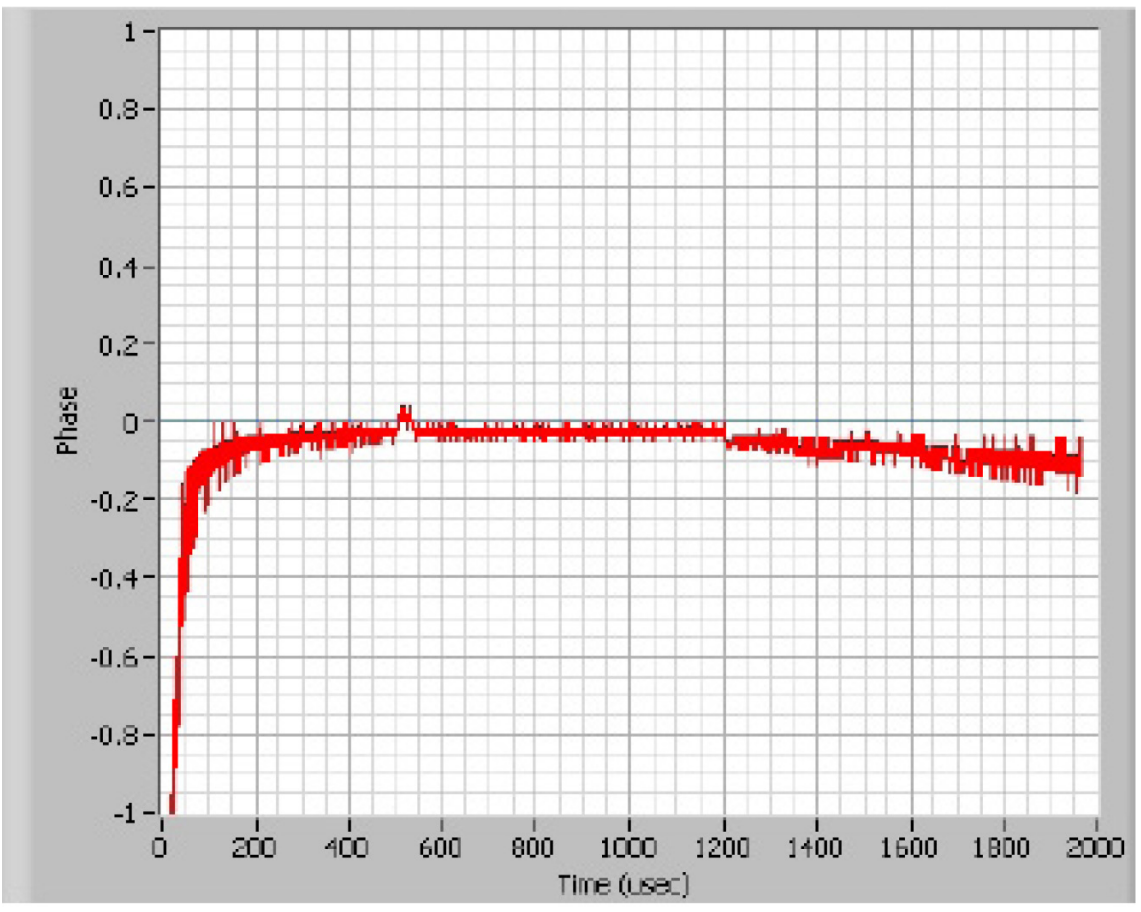

Figure 5: Vector Sum Phase

## System Performance

The operation of the LLRF system in closed loop was tested using the cavity simulator with a 200 Hz half bandwidth for the cavity. The receiver/transmitter was included in the closed loop along with an RF amplifier on the drive signal and an 8-way splitter to provide the cavity probe signals. The feed-forward signal was used to drive the loop with a set-point table adjusted for a cavity τ of 700µs and an amplitude to provide an error of about 10 percent in amplitude and 10 degrees of phase error. The feedback loop was closed with a proportional gain of ~450 and an integral gain of 2e7 rad/s with a pole of 300 Hz.

The amplitude and phase regulation of the vector sum are shown in Fig. 4 and Fig. 5. The amplitude error was about .01% rms and the phase error was about 0.01° rms. This compares well with the results obtained with a single cavity operation at Capture Cavity II [3].

## References

[1] P. Varghese et al., “Multichannel Vector Field Control Module for LLRF Control of Superconducting Cavities”, PAC’07, Alburquerque, June 2007.

[2] U. Mavric et al., “A 96 Channel Receiver for the ILCTA LLRF System at Fermilab”, PAC’07, Alburquerque, June 2007, WEPMN102.

[3] E. Harms et al., “Operating experience with CC2 at Fermilab’s SRF Beam Test Facility”, LINAC 2010, Tsukuba, Sep 2010, THP029.